\documentclass[10pt,preprint]{aastex}

\def\a0{a_0}
\def\msun{M_{\odot}}

\def\cmst{{\rm cm~ s}^{-2}~}

\def\beq{\begin{equation}}
\def\eeqno#1{\label{#1}\end{equation}}

\def\msun{M_{\odot}}
\def\az{a_{0}}

\def\l0{\ell_{0}}

\def\l{\lambda}

\def\a{\alpha}

\def\az{a_{0}}

\usepackage{natbib}
\usepackage{amssymb}
\usepackage{epsfig}
\begin{document}

\title{MOND: time for a change of mind?}

\author{Mordehai Milgrom}
\affil{Center for Astrophysics, Weizmann Institute, Rehovot 76100,
Israel}

\section{What is MOND?}
\label{section1} MOND is a quarter-century old paradigm of dynamics
propounded as replacement for Newtonian and relativistic dynamics.
It greatly departs from these venerated theories when accelerations
in a system under study are very small: much smaller than those
encountered on earth or in the solar system, but quite typical of
galaxies and other galactic systems. I shall limit the discussion to
non-relativistic MOND, applicable for most problems now under study
in galactic astronomy. MOND introduces into physics a new constant
$\az$, with the dimensions of acceleration, which  marks the
boundary of validity of classical dynamics, and also features
prominently in the phenomenology of low acceleration systems. In
this sense, the role of $\az$ is similar to that of the speed of
light, $c$, in the context of the relativistic departure from
classical physics, or the role of the Planck constant, $h$, in the
quantum context. It appears in the equations of dynamics of any
theory that is anchored in the MOND paradigm, but for systems with
accelerations that are much larger than $\az$, such a theory reduces
to the standard, Newtonian theory. This can be achieved, formally,
by simply substituting everywhere in the MOND equations $\az=0$,
which should reduce them to the equations of standard physics (just
as taking an infinite speed of light, or $h=0$, reduces relativistic
results, or quantum results, to the corresponding classical limit).
\par
But how does the theory look like and what does it imply for
accelerations that are smaller than $\az$?  The most appealing way
to describe MOND for very low accelerations is to point to a
symmetry principle: in this limit the theory becomes invariant to
simultaneous scaling of times and distances: Take any physically
permissible configuration--such as a planetary system, or a galaxy
of stars--and increase all distances in the system by a certain
factor, and also increase all times by the same factor. The symmetry
implies that we then obtain another physically permissible
configuration. For example, in a planetary system that has very low
accelerations and obeys the MOND dynamics, a result of the symmetry
is that the length of the planetary year increases in proportion to
the orbital radius, and the orbital velocities are then independent
of the orbital radius. In contrast, standard dynamics do not have
this symmetry: if we increase the radius of the earth's orbit around
the sun by a factor of four, say, the orbital time will increase by
a factor of eight, not by a factor four as would be required by such
a symmetry.
\par
More generally, the symmetry dictates how the orbital (centripetal)
acceleration, $a$, of a body on a circular orbit of radius $R$
around a central mass, $M$, depends on these parameters in the very
law acceleration limit: Since $a$ can depend only on $R$, $M$, and
on the constants $G$ and $\az$, the symmetry and dimensional
considerations tell us that $a$ must be proportional to
$\sqrt{MG\az}/R$, and we define the value of $\az$ once and for all
so as to have equality  $a=\sqrt{MG\az}/R$. This dependence is in
sharp contrast with the Newtonian relation $a=MG/R^2$. The main
differences to note are: (i) the MOND acceleration decreases in
inverse proportion to $R$, not to $R^2$, and (ii) it is no longer
proportional to the attracting mass, but to its square root. Note
also that MOND generically gives higher accelerations than Newtonian
dynamics, for a given mass.
\par
\begin{figure}
\centerline{\psfig{figure=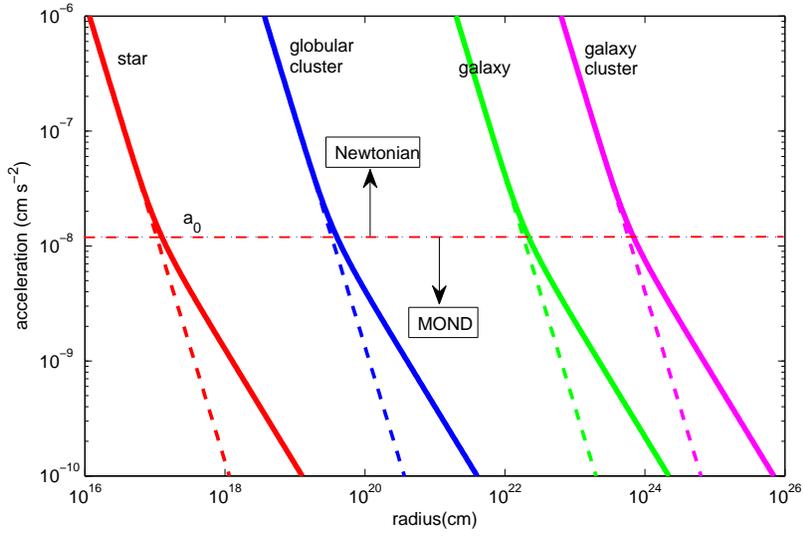,width=12cm}} \caption{The
MOND centrifugal acceleration of a particle on a circular orbit
around a mass $M$, as a function of orbital radius (in heavy lines),
for a star of one solar mass ($1\msun$)(red), globular cluster of
mass $10^5\msun$ (blue), a galaxy of mass $3\times 10^{10}\msun$
(green), and a galaxy cluster of mass $3\times 10^{13}\msun$
(magenta). The Newtonian accelerations are shown as dashed lines.
Departure of MOND from Newtonian dynamics occurs at different radii
for different central masses, but always at the same value of the
acceleration, $\az$, below which we are in the MOND regime, and
above which we are in the Newtonian regime.} \label{pointmass}
\end{figure}

Any theory based on the MOND paradigm should interpolate between the
classical and the MOND limits and should thus cover the full range
of phenomena from very low to very high accelerations. For example,
if we have a planetary system with high accelerations for the inner
planets and accelerations much smaller than $\az$ for the outer
ones, the theory should describe the system in full, accounting for
both limits correctly. In such a theory the dependence of the
acceleration on mass and radius will be of the form
$a\mu(a/\az)=MG/R^2$, such that at small radii, where $a$ is much
larger than $\az$, $\mu$ is very nearly unity, so the Newtonian
expression is valid, and $\az$ disappears. At large radii, where $a$
is small compared with $\az$, we have to have $\mu(x)=x$ to good
accuracy, and the MOND relation obtains. Figure \ref{pointmass}
shows the schematic dependence of $a$ on $M$ and $R$, compared with
the Newtonian prediction, for a range of masses representing various
astronomical objects.

The function $\mu(x)$ encapsules the transition from the classical
to the modified regime. Such functions appear commonly in other
instances of modifications of classical physics. For example,
Planck's black-body function describes the spectrum of the black
body with full classical-quantum coverage. It reduces to the
classically predicted spectrum at low frequencies (where the Planck
constant does not appear). The Lorentz factor, which appears in
various expressions of special relativistic kinematics and dynamics,
is another famous interpolating function between the classical and
relativistic regime.

\section{Why MOND? The mass discrepancy and the appearance of dark matter}
Why consider such a major modification to Newtonian dynamics and
General Relativity when it is common knowledge that these two are so
very successful in accounting for non-quantum phenomena? The truth
of the matter is, however, that these theories fail miserably in
accounting for the observed dynamics of most galactic and
cosmological systems--such as galaxies, binary galaxies, small
groups of galaxies, and rich galaxy clusters. These theories can be
saved only if new, ad hoc, dominant ingredients of matter-energy are
introduced into the universe; these ingredients are known as ``dark
matter'' (DM) and ``dark energy'' (DE). Standard dynamics fail to
explain the observed motions in such systems if the only mass
present is what we directly observe (stars, cool and hot gas, etc.).
The velocities in these systems are so high that, with only the
observed mass to arrest them gravitationally, the systems would just
fly apart in a relatively short time. The amount of extra mass
needed to keep such systems from dispersing varies from object to
object, and from place to place within an object, but it is
routinely found to be several times to tens of times larger than the
mass observed directly. The dark matter paradigm then assumes that
whatever extra mass is needed is there in the form of ``dark
matter'', of some yet unspecified nature.

\begin{figure}
\centerline{\psfig{figure=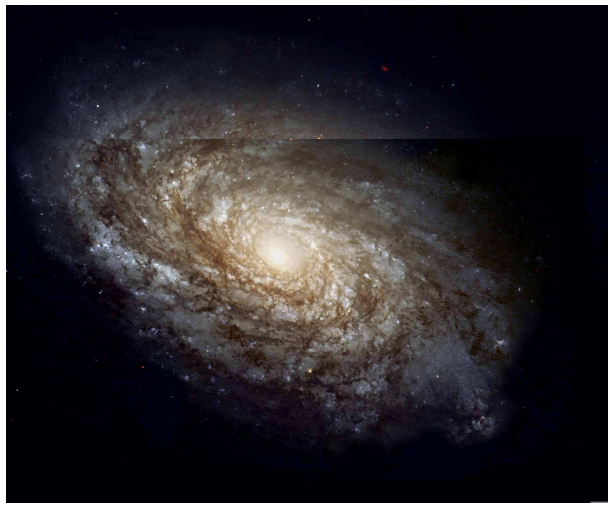,width=8cm}} \caption{A
spiral (disc) galaxy.} \label{disc}
\end{figure}

\begin{figure}
\centerline{\psfig{figure=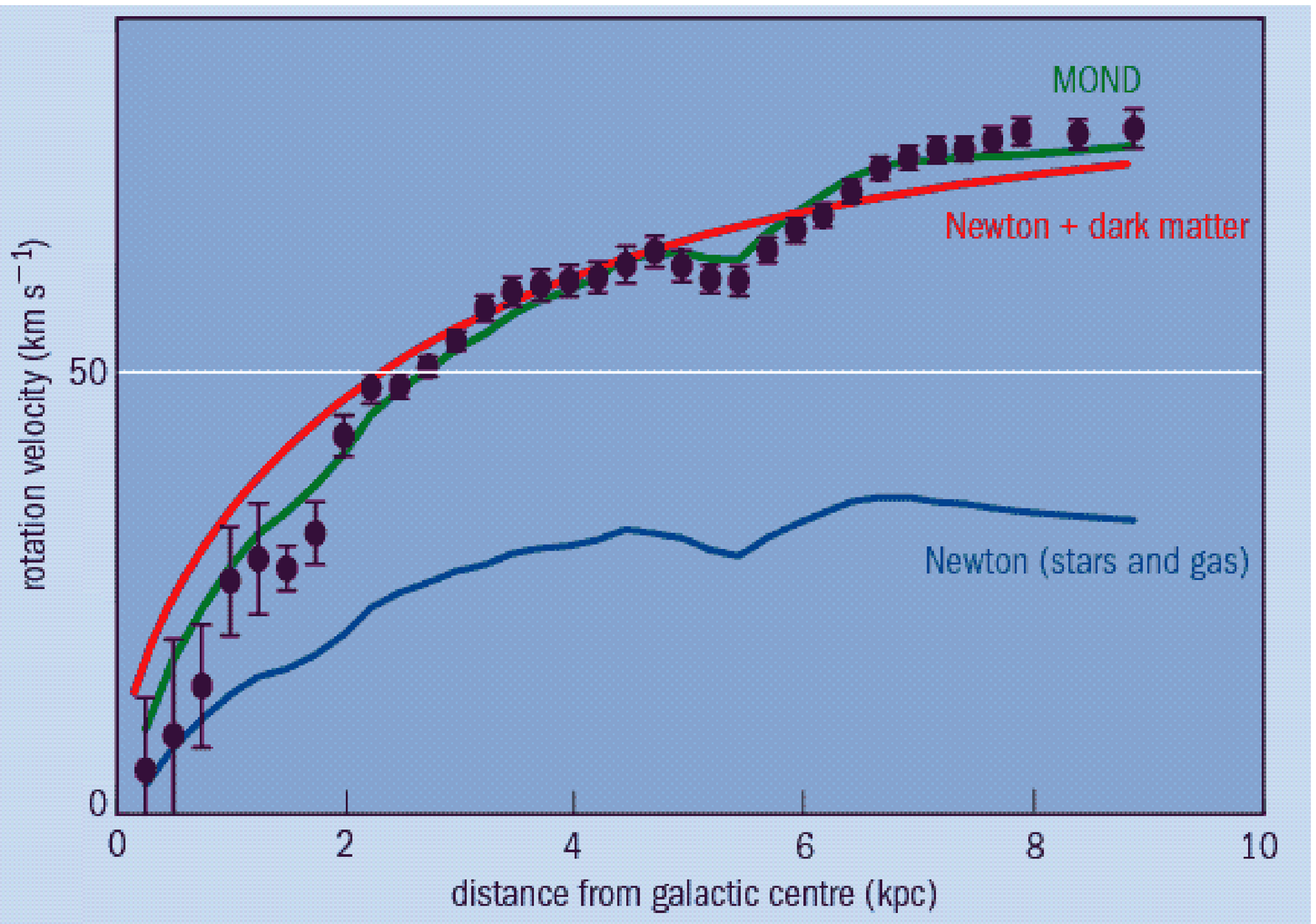,width=10cm}} \caption{The
measured rotation curve of the galaxy NGC1560 shown by the data
points. The predicted Newtonian curve based on the measured mass
distribution is shown in blue. It shows a velocity disparity of a
factor 2.25 at the last measured point, which corresponds to a
factor-of-five mass discrepancy. The MOND prediction is shown in
green. The best fit with a dark matter hallo of the type predicted
by CDM simulations is shown in red. I has two free parameters over
which the fit is optimized: the mass and scale of the halo (no such
freedom exists for MOND).  Courtesy of Stacy McGaugh.} \label{1560}
\end{figure}

Consider just one example of such measurements leading to a mass
discrepancy, arguably the most reliable and clear-cut to date. A
disc galaxy, such as is shown in Fig.\ref{disc}, is made of gas and
stars in a rather thin, flat disc; both components circle the galaxy
on nearly circular orbits with a velocity $V$ that depends on the
radius, $R$, of the orbit. The examination of the galaxy rests on
two kinds of measurements: First, we measure the dependence of $V$
on $R$ to give the so-called rotation curve, $V(R)$, as is depicted
in Fig.\ref{1560}. Second, we measure the distribution of mass we
see in the galaxy in the form of stars and gas. Given the mass
distribution, one calculates the gravitational field of the galaxy
using one's favorite theory of gravity (e.g., Newtonian gravity or
MOND). From this one calculates the expected orbital velocities
based on the law of inertia by which we equate the centripetal
acceleration $V^2(R)/R$ to the gravitational force per unit mass. It
is a fact established for {\it all} disc galaxies studied to date
that the Newtonian prediction of the rotational speeds fall short of
the measured ones. In some galaxies the discrepancy appears only
beyond a certain radius, and then increases with radius; in many
other cases the discrepancy is present at all radii. For the
rotation curve shown in Fig.\ref{1560} the discrepancy exhibited by
the predicted Newtonian velocities is of a factor 2.25 in the outer
parts, corresponding to a factor of five discrepancy in mass. So, in
this case a dark matter mass four times as large as that of normal
matter is needed. Similar analyses lead to similar mass
discrepancies in all astronomical systems from the scale of galaxies
and up.

Such mass discrepancies are deduced not only by following the motion
of massive objects bound to the system such as stars and gas. A
powerful method that complements such studies involves the
measurement of the bending of light rays coming from far objects in
the background, as they pass near the object under study--so called
gravitational lensing. The results of such measurements are
consistent with those of bound particles measurements in regions
where the two coincide. The method also enables us to measure the
discrepancy at very large radii, where the first method is not
applicable for lack of probing objects.

The need for DM arises also in the context of cosmology, in
particular in connection with the way galactic systems have formed
out of the uniform, primordial soup. It is thought that the
universe, which has been expanding since the initial ``big bang'',
contained in the early stages a rather uniform, hot mixture of known
matter (protons, electrons, photons, some nuclei, etc.) and DM in
roughly a one to five proportion. As the universe expanded and
cooled, the small seed non-uniformities in the mass distribution
have gradually increased in magnitude, due to gravitational self
attraction, a process that eventually led to the formation of
galactic systems. Normal matter was initially hot and ionized
(charged) and so submissive to the black body radiation ambiance
then present in abundance. In such a state, the normal matter could
not efficiently collapse to enhance the strength of the mass
agglomerates, because the domineering radiation does not collapse.
However, at a certain stage, matter cooled enough and formed neutral
atoms. Its non-uniformities--now free of the restraints of the
radiation--could then continue to grow effectively.

Alas, without auxiliaries, standard dynamics tells us that such
non-uniformities of normal matter did not have enough time to grow
into the structures we see today in the time since it became
neutral. Dark matter comes to the rescue because, being neutral, it
could have collapsed freely even before the neutralization. When
normal matter neutralized it already found itself in the presence of
better developed clumps of DM, which then pulled the normal matter
into them, hastening the collapse. Thus begins a complex process of
continued collapse, interaction and mergers between clumps,
dissipation of the normal-matter gas, its forming stars, and its
expulsion from galaxies by various processes, etc.. This, still
ongoing, process has led to the matter agglomerates that we see
today as galaxies, galaxy groups, clusters, and super-clusters.
These agglomerates are composed of an extended invisible ``halo'' of
DM, at the center of which sits the normal matter, which is visible
as radiation emitting and absorbing gas, and as shining (and dead)
stars that formed later.  All this is what the DM paradigm would
have us believe. (Dark energy will be discussed later.)

The DM paradigm can make hardly any prediction without further
specification of the nature of the DM. As a general concept DM is
only a filler, whose presence is assume where needed. However, there
is a class of candidates for the  DM substance that is now favored,
called cold DM (CDM). Attention on CDM has converged after several
other candidates have been ruled out and discarded over the years,
such as neutrinos, and massive, dead, stellar objects. This
particular choice does lend itself to certain predictions, and from
now on I shall refer to this option.

There are, at present, two  types of particles that are
considered the leading candidates for the CDM constituents: supersymmetric counterpart of
ordinary particles is
the one, the so called ``axion'' is the other. Both are hypothetical particles
whose existence is rooted in  different elementary particle
theories. They are of very different nature, so the ways to detect
them are very different, and the ways they have appeared in the
cosmos, in the first place, are very different. However, their
effects in cosmology are similar, as they simply act as inert,
hardly interacting, relatively heavy particles.

Neither of the two has been produced in the laboratory, or is even
known to exist on other secure grounds, let alone to have been
caught in the act of playing the role of DM (even if such particles
do exist, they need not be the DM particles). One of the declared
goals of the LHC accelerator in CERN is to produce and detect the
supersymmetric partners in the debris of high-energy subatomic
collisions. If these are indeed discovered it will tell us at least
that a candidate exist. But, to establish it as a DM particle we
need to detect these particles directly as dark matter: The solar
system, and with it earth, is presumed to be bathed in the sea of
dark matter particles engulfing our galaxy, as it does all galaxies;
these could, in principle, be detected directly. There is a number
of underground experiments afoot, trying for some years now to do
just that.

Having taken all that in, it has to be realized that the deduction
of large mass discrepancies is based on a combination of the law of
gravity (the way gravity depends on mass and distance), and on the
law of inertia (describing how motion responds to an applied force).
These are the central pillars of Newtonian dynamics, and hence also
of General Relativity, which rests on Newtonian dynamics. They work
very well in the laboratory and the solar system; but, can they also
be applied in the realm of the galaxies?

Enters MOND: it is possible to device a new theory of
dynamic--incorporating standard dynamics at high accelerations, but
forgoing it at low ones--that explains almost all aspects of the
mass discrepancies in galactic systems with no need to invoke dark
matter. This is what MOND claims to achieve.

And the ``dark energy''? MOND does not obviate it; in fact, as we
shall see below, MOND may be part and parcel of a universe governed
by ``dark energy''.

\section{What does MOND predict and how does it perform?}
\begin{figure}
\centerline{\psfig{figure=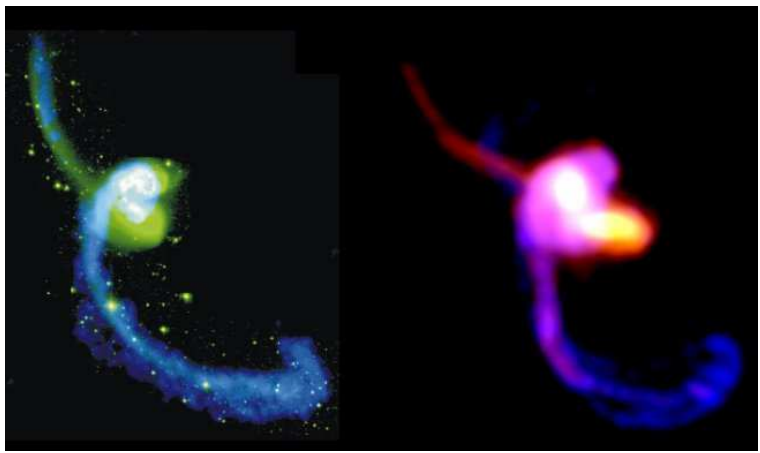,width=10cm}} \caption{
Simulation of closely interacting galaxy pair known as The Antennae
with MOND (right) compared to the observations from Hibbard et al.
2001 (left). In the observations, the gas is represented in blue and
the stars in green. In the simulation the gas is in blue and the
stars are in yellow/red. Considering that many of the details of the
history of the collision are not known and cannot be incorporated
in the simulation, the agreement is quite remarkable. From Tiret and
Combes 2008.} \label{tiret}
\end{figure}

One can construct various detailed theories that incorporate the
basic MOND tenets listed above. For example, Newtonian gravity is
known to be governed by the Poisson equation, by which the
gravitational potential is determined from the mass distribution
(the same equation governs the electrostatic field). This equation
has been generalized to incorporate the MOND tenets. The resulting
MOND theory is as complete a theory as Newtonian dynamics, and has
been the basis for many analytic and computational studies. A recent
example of numerical simulations based on this theory of the
complex, interacting binary galaxy system known as The Antennae, by
Tiret and Combes from the Paris Observatory, is shown in
Fig.\ref{tiret}.

There are other theories that one can write for the nonrelativistic
regime. In addition, we would like to construct a MOND theory that
is compatible with the principles of relativity. There has been
progress made in this direction, notably the relativistic MOND
formulation call TeVeS proposed by Jacob Bekenstein from the Hebrew
University, but it is felt that even this is not the ultimate
theory.

Luckily, there are many predictions of MOND that follow directly
from the basic premises and do not require a specific theory. These
may be viewed as the MOND analogues and extensions of Kepler's laws
for planetary motions. Kepler discovered his laws as purely
phenomenological regularities in the motions of the planets in our
solar system, without understanding their origin. These laws then
served Newton in deriving his general theory of dynamics, which not
only reproduced Kepler's laws, but have greatly extended them. For
example, Newton's theory also predicted the motion of unbound bodies
on hyperbolic orbits, and of planetary motions in other systems,
around other central stars. More generally, it predicted the
behavior of arbitrary systems governed by gravity.

\begin{figure}
\centerline{\psfig{figure=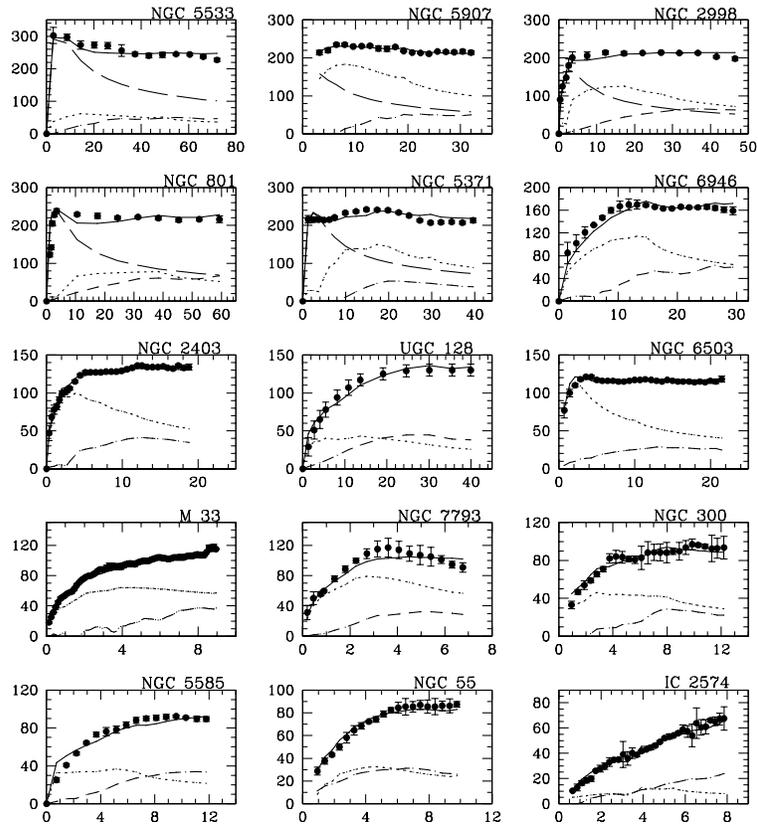,width=12cm}} \caption{The MOND
rotation curves for several galaxies, shown as the solid lines going
through the data points (the other lines show Newtonian curves
calculated for  different mass components such as stars and gas).}
\label{rcs}
\end{figure}

A  MOND theory, likewise, predicts the general behavior of arbitrary
systems held by gravity. As in the Newtonian case, such predictions
are usually deduced from involved calculations (for instance,
solving the motions of the planets in our solar system in Newtonian
dynamics, with all the interactions between planets taken into
account, is a formidable, yet incomplete task). But, in MOND too,
one can extract a set of general regularities that can be directly
applied to galactic systems without further calculations. Here are
some examples:

1. As we saw already, MOND predicts that orbital velocities on
circular orbits around a concentrated mass become independent of the
orbital radius, for large radii (where the acceleration becomes
smaller than $\az$). For disc galaxies, this means that the
rotational velocity should become constant with radius at large
radii, as indeed it does (see Fig.\ref{rcs}).

2. We also saw that the constant asymptotic
rotational velocity in a galaxy should be proportional to the fourth
root of the galaxy's mass. This is also in very good agreement with observations.

3. As summarized in Fig.\ref{pointmass}, the mass
discrepancy in galaxies should appear at a different
radii for different galaxies, but always at the same
value of the centrifugal acceleration $V^2/R=\az$. Galaxies for
which the acceleration is smaller than $\az$ at all radii, should
show a discrepancy everywhere. All this is amply born out by the
observations.

4. MOND also predicts that an inflated, spheroid-like halo of dark
matter, as is predicted by the cold dark matter paradigm, should not
suffice to explain all the facets of the mass discrepancy in disc
galaxy: an additional flat, disc-like component, with predictable
properties, should be necessary.

5. MOND predicts that the discrepant acceleration in galactic
systems can never much exceed $\az$.

Quite a few more such predicted laws are known (e.g., ones
pertaining to elliptical galaxies and other such systems). They all
conform well with the data.

Above and beyond such Kepler-like laws, the flagship of MOND
phenomenology is the full prediction of the exact rotation curves of
individual disc galaxies: MOND does this for each and every galaxy,
given only the observed distribution of the normal mass in the
galaxy. Over a hundred galaxies have been analyzed in this way to
date, with generally, very good success. We already saw one example
shown in detail in Fig.\ref{1560};  some more are shown in
Fig.\ref{rcs}.

\section{ MOND vs. CDM}
The fact that the initial motivation for DM and MOND is similar, and
that they aim to account for similar phenomena, may give the
impression that they are similar paradigms that can be tested by
similar criteria. This is anything but true: MOND is much more
predictive, and lends itself to falsification to a far greater
degree then the CDM paradigm, to say nothing of DM in general. MOND
makes definite predictions on dynamics for {\it individual} objects
based on just the observed mass distribution. For CDM, predictions
of this kind are impossible, with very few exceptions, as they would have to hinge on understanding of the interrelations between normal and dark matter in a given system, which, in turn, would
result from complex formation and evolution histories that cannot be
known for a given object. This is further complicated by the fact
that normal matter and DM follow very different evolution paths.
\par
CDM is capable of making some predictions on general properties of
the bare CDM halos themselves, if we neglect the effects of normal
matter on the CDM: Using high power computer simulations, a uniform
distribution of CDM, seeded with some initial irregularities, can be
evolved to obtain some {\it statistical} properties of CDM halos at
the present time. This includes population statistics, such as the
distribution of total masses of these halos, and general intrinsic
properties of the halos such as the form of the density distribution
in them. However CDM is almost completely dumb on the expected
properties of the normal matter inside these halos, and on
interrelations between CDM and normal matter; why? Normal matter is
far more active and interactive than the purported CDM; it undergoes
more complex and erratic influences: it can emit and absorb
radiation, it is subject to energy losses by dissipation, it
interacts with magnetic fields, it can be ejected from galaxies by
stellar explosions, it can form stars, etc., etc., all of which the
CDM does not partake. It is enough to glance at galaxies to see that
the normal matter in then have very different characteristics from
those attributed to CDM, even though they are presumed to have
started as a well mixed soup: In many galaxies the normal matter
forms a thin rotating disc, while the DM is thought to constitute a
spherical or elliptical halo that is hardly rotating. The DM halo is
much more extended then the normal matter in galaxies. The ratio of
DM to normal matter in galaxies is much larger than the cosmic ratio
with which they presumably started. All this tells us quite cogently
that the two components should not be well correlated, and that
their mutual relations are unpredictable in the CDM paradigm. Even
the most basic fact one would like to be able to predict: the total
amount of CDM to be expected around a given visible galaxy, cannot
be estimated, to say nothing of finer details, such as the exact
distribution of DM around a given, visible galaxy.

MOND, on the other hand, is capable of predicting all this (or
rather the equivalent, since it repudiates DM).

Take, as an example, the rotation curve shown in Fig. \ref{1560};
it demonstrates clearly the differences between the predictive
power and success of MOND and CDM on galactic scales. The figure shows the measured curve together with the
MOND curve, which is {\it predicted} with hardly any
freedom. MOND does well in predicting the general shape of the
rotation curve, and even reproduces the peculiar feature consisting
of a ``dip''. In contrast, the CDM curve, which is also shown, is not a prediction, as it
involves also a dominant contribution from a DM halo with a priori
unknown properties. Numerical simulations yield CDM halos that have
a certain form of the density distribution characterized by two
structural parameters: the mass of the halo and its characteristic
radius (as well as some unpredictable degree of departure from
sphericity; but, in rotation curve fits the halo is usually taken as
spherical). The observed velocity curve is then fitted with such a
halo optimizing the fit with a necessary best choice of mass and
size for the halo. This allows  great freedom in fitting to the
data, as any quantity of DM can be assumed with impunity. And yet,
even with all its freedom, the best CDM fit in Fig. \ref{1560} is
rather inferior. It misses the general trend, and totally fails to
reproduce the observed ``dip''. (The dip results from a feature in
the mass distribution in the gas disc, and the putative spherical
halo ``erases'' the effect.)

\begin{figure}
\centerline{\psfig{figure=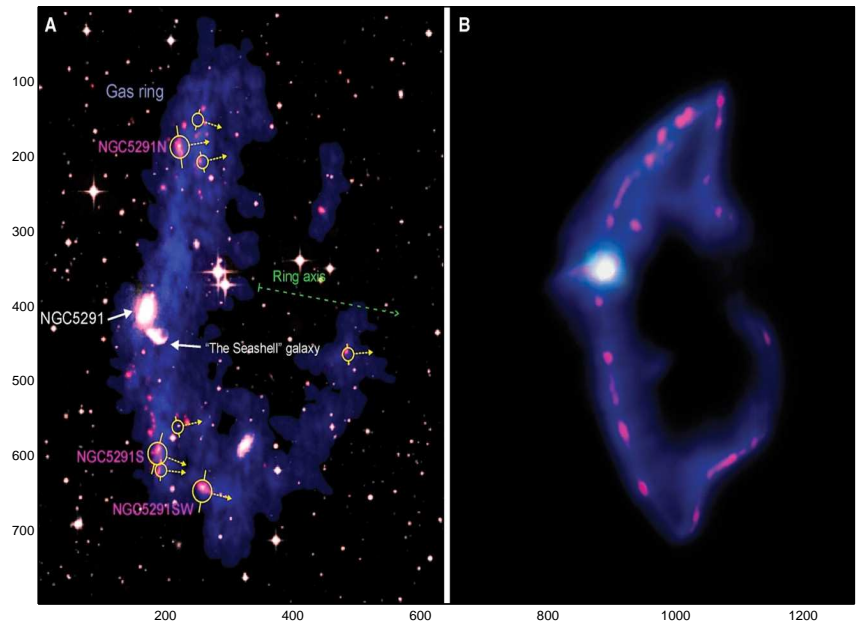,width=14cm}} \caption{Left: The
gas ring ejected from the high speed collision of the galaxy NGC5291
with another galaxy (present position outside of frame). The forming
phoenix dwarfs NGC5291N,S,SW are marked. Right: Numerical simulation
of the collision and its outcome. Curtesy of Frederic Bournaud.}
\label{debris}
\end{figure}

On rare occasions the CDM paradigm does predict the amount of dark
matter that should accompany a normal matter galaxy. A case in point
involves the dwarf galaxies that form out of the debris of
high-velocity collisions between cut-and-dried disc galaxies. The
colliding galaxies come with their purported halos of CDM and
collide violently, ejecting out some of their mass (both normal and
DM) and continuing on their speedy way. The ejected debris typically
form semi-ordered, elongated structures of gas, such as rings or
arcs, coming from the colliding gas discs. An example of such a
ring-like debris is shown in Fig.\ref{debris}. (The interacting
galaxies shown in Fig.\ref{tiret} are believed to undergo a milder,
lower-speed collision and will eventually merge; they too eject
``antennae'' of gas and stars.) From the gas ashes of the debris,
little phoenix galaxies form in the course of time. In the system
shown in Fig.\ref{debris}, believed to have resulted from a
collision some half a billion years ago, a few such little phoenixes
have been identified and studied in detail. The CDM, which
supposedly come from the spheroidal halos of the colliding galaxies,
is spread all around with hardly any of it accompanying the gas, and
hardly any finding its way into the phoenix dwarfs. This assertion
is strongly supported by numerous simulations (the right panel in
Fig.\ref{debris} shows the gas structure resulting from such a
simulation of the system on the left). In a unique instance of
predictability on normal-to-dark-matter relation, the CDM paradigm
thus predicts no, or very small, mass discrepancies in these
phoenixes. And yet, analysis of the three accessible phoenixes in
the system found all three to show mass discrepancies of about a
factor three, in conflict with the prediction of the CDM paradigm.
\par
And what does MOND predict for these? These dwarfs are measured to
rotate with centripetal acceleration a few times smaller than $\az$,
and MOND thus predicts substantial mass discrepancies in them, just
at levels and distributions that match those found.

The DM paradigm is known to face a number of other severe
difficulties in explaining the observed properties of galaxies,
which I shall not discuss here, and which DM advocates are hard at
work to explain away.

\section{Deeper significance?}
MOND  already boasts two main achievements: it obviates the need for
dark matter in galactic systems, and it predicts, explains, and
organizes a large body of galactic dynamics data. But, MOND's most
far reaching ramifications may yet be in store. This expectation is
based on the following observation: The MOND acceleration constant,
$\az$, appears in quite a few predictions of MOND regarding dynamics
of galaxies (just as the Planck constant appears in the description
of disparate quantum phenomena, such as the black body spectrum,
atomic spectra, superconductivity, the quantum Hall effect, etc.).
Comparing these predictions with the observations, we can determine
its value in several independent ways. The value obtained is the
same for all methods, within the uncertainties, and is $\az\approx
1.2\times 10^{-8}\cmst$. This value is tantalizingly close to
acceleration constants that typify our universe as a whole. One such
cosmic acceleration constant, call it $a_H$, is gotten by
multiplying the speed of light by the expansion rate of the
universe, known as the Hubble constant. (To envisage this numerical
proximity, note that a body starting from rest and accelerating with
acceleration $\az$ will approach the speed of light in the life time
of the universe since the big bang.) Another constant, call it
$a_{\Lambda}$, is derived from the recently observed acceleration of
the expansion rate of the universe. This accelerated expansion is
difficult to explain in the context of standard physics (General
Relativity) if only matter of conventional properties is present
(normal or dark matter). The majority view is that the acceleration
is due to the dominant presence in the cosmos of an entity known as
``dark energy'', whose gravity acts, unlike that of conventional
matter, to accelerate the expansion; it is required to make up about
75 percent of the mass-energy content of the cosmos today. In
General Relativity, gravity and geometry of space-time are one and
the same. The gravity effect of the ``dark energy'' also means that
our space-time is approximately a ``spherical'' one with a radius
that is related to the acceleration rate. It is a great mystery that
for some reason the values of the seemingly unrelated constants
$a_H$ and $a_{\Lambda}$ are today nearly equal ($a_H$ varies over
time, while $a_{\Lambda}$ could be a veritable constant). It is yet
another mystery that $\az$, which apparently derives from completely
unrelated phenomena, is also numerically close to these two.

Is this a mere coincidence? Personally I view it as a strong hint
that MOND, which pertains to objects that are very small on
cosmological scales is, nonetheless, strongly related to the global
state of the cosmos. Exactly how? we do  not yet know for sure, but
there are ideas in this vein. The hope is that when the connection
is understood, MOND, and the value of $\az$ will follow from this
understanding. The situation might be schematically similar to the
dynamics of moving bodies near the earth surface, which was found to
be governed by some acceleration constant $g$, Galilei's free-fall
acceleration. As long as we only perform experiment near the earth's
surface we appear to have a theory with some acceleration constant.
But once we understand gravity better, we see that this is only an
approximate description, and that the constant can be derived from
properties of the earth, such as its mass and radius.

One interesting consequence of the above numerical proximity of
$\az$ to cosmic accelerations is that there cannot exist in nature
systems with both relativistic gravity and deep-MOND accelerations
(such as a black hole with low accelerations): such a systems would
have to be much larger than the visible universe.

The MOND-cosmology connection is further supported by the
identification of symmetries that are common to the deep MOND limit,
and a dark-energy dominated space-time.

\section{Open questions}
MOND is a paradigm still under construction. In the nonrelativistic
regime we have a working, full fledged theory, as described above,
but we are not sure that it is the ultimate one, since there are
other possibilities. We also have a working relativistic formulation
of MOND called TeVeS. This replaces Einstein's field equations for
calculating the geometry of space-time, or the gravitational field,
in a given system. TeVeS is already a great advance, because it
enables us to calculate the effects of gravitation on light rays
(gravitational lensing), and the cosmological evolution of the
universe, which are beyond the capabilities of nonrelativistic
theories. However, TeVeS too seems to have its limitations, and the
quest for a relativistic formulation of MOND continues. In any
event, theories such as TeVeS do not embody the above mentioned
connection of MOND with cosmology; one may then wish for a more
fundamental theory that does.

On the phenomenological front MOND is also not all roses. For
example, the need for dark matter in the cosmological setting has
not yet been convincingly  explained, although there have been
suggestions on how this too may be replaced by a MOND effect. Also,
MOND does not completely explain away the need for dark matter in
galaxy clusters: MOND does reduce substantially the amount of DM
needed, but there is a remaining discrepancy between the required
and the directly observed masses in clusters. We, MOND advocates,
attribute this to normal matter that could easily have hidden in
clusters, perhaps in the form of massive neutrinos, perhaps as
extinguished stars, or in the form of cool, dense gas clouds. Note
that, in any event, even the normal matter believed to exist in the
universe is still largely unaccounted for, and the amount still
needed in clusters makes up only a very small fraction of the normal
matter that is still missing.

All in all, these are exciting times for those involved in MOND
research, with a tailwind of continuing successes, and the
attraction of remaining challenges.

\section{Further reading:}
\begin{itemize}
\item
Adam Frank: The Einstein Dilemma, ``Discover'' Magazine, August 2006

(http://discovermagazine.com/2006/aug/cover/?searchterm=milgrom)
\item
Marcus Chown: Dark Matter is Dead, ``BBC Sky at Night'' Magazine,
September 2007

(http://www.skyatnightmagazine.com/viewIssue.asp?id=837)
\item
The MOND pages (http://www.astro.umd.edu/~ssm/mond/index.html).
Various accounts of MOND with a link to an extensive list of papers
on MOND
\item
M. Milgrom: Does Dark Matter Really Exist? Scientific American,
August 2002 (appeared in Hebrew as well).
\end{itemize}
\end{document}